\documentclass[aps, prl,nofootinbib,tightenlines, twocolumn,floatfix]{revtex4}

 \usepackage[dvips,final]{graphicx}
  \usepackage{amssymb}
   \usepackage{amsmath}
    \usepackage{amsfonts}
     \usepackage{epsfig}
      \usepackage{bm} % bold math
\usepackage{mathrsfs}
\usepackage{slashed}

\usepackage[colorlinks = true,
linkcolor = blue,
urlcolor  = blue,
citecolor = blue,
anchorcolor = blue]{hyperref}

\begin{document}

\title{Hydrodynamic manifestations of gravitational chiral anomaly}

\author{G. Yu. Prokhorov}
\email{prokhorov@theor.jinr.ru}
\affiliation{Joint Institute for Nuclear Research, Joliot-Curie str. 6, Dubna 141980, Russia}
\affiliation{Institute of Theoretical and Experimental Physics, NRC Kurchatov Institute,
B. Cheremushkinskaya 25, Moscow 117218, Russia}
\author{O. V. Teryaev}
\email{teryaev@jinr.ru}
\affiliation{Joint Institute for Nuclear Research, Joliot-Curie str. 6, Dubna 141980, Russia}
\affiliation{Institute of Theoretical and Experimental Physics, NRC Kurchatov Institute,
	B. Cheremushkinskaya 25, Moscow 117218, Russia}
\author{V. I. Zakharov}
\email{vzakharov@itep.ru}
\affiliation{Institute of Theoretical and Experimental Physics, NRC Kurchatov Institute,
B. Cheremushkinskaya 25, Moscow 117218, Russia}
\affiliation{Pacific Quantum Center,
Far Eastern Federal University, 10 Ajax Bay, Russky Island, Vladivostok 690950, Russia\vspace{0.8 cm}}

\begin{abstract}
The conservation of an axial current modified by the gravitational chiral anomaly implies the universal transport phenomenon (Kinematical Vortical Effect) dependent solely on medium vorticity and acceleration but not dependent explicitly on its temperature and density. This general analysis is verified for the case of massless fermions with spin 1/2.
\end{abstract}

\maketitle

%================
\section{Introduction}
\label{sec_intro}
%================

Discovery of the quark-gluon plasma, with its unusual properties,
changed the landscape of theoretical disciplines, focusing attention
on theory of fluids, quantum fluids in particular, see e.g.
\cite{Schafer:2009dj, Brauner:2022rvf}. While
the low value ratio of viscosity to entropy density $ \eta/s $ of the plasma
and fast set of equilibrium are still awaiting their explanation, theory
of quantum chiral effects, in particular, chiral magnetic effect (CME) $ {\bf j} \sim {\bf B}$, where $ {\bf j} $ is the vector current and $ {\bf B} $ is the magnetic field, has been developed
to the point which allowed for a massive experimental effort  to establish
(or reject) it \cite{Kharzeev:2022hqz}.
Experimentation with the quark-gluon plasma is all the more
suited for study of chiral effects since the observability of these effects is closely
related to the approximation of fluid being ideal \footnote{Initially, chiral effects were obtained using just the zero viscosity approximation (e.g. \cite{Son:2009tf}), although one should expect them to be more universal.}.

Study of heavy-ion collisions could also provide with a window
to observe imitation of gravitational effects \cite{Castorina:2007eb, Kharzeev:2005iz}. Indeed, the STAR collaboration
concluded that the properties of the quark-gluon plasma favor the models
which assume the plasma being produced in a rotated and accelerated state \cite{STAR:2017ckg}. A dual
description of kinematics of acceleration and rotation in terms of gravitational
potentials goes back to Einsteins lectures of the general relativity \cite{einstein}.  In the context
of transport phenomena the similarity between the gravitational and entropic forces has been
emphasized by Luttinger \cite{Luttinger:1964zz}. The most advanced suggestion in this direction is the hypothesis
that the fundamental gravitational interaction can be replaced by its thermodynamic counterpart \cite{Verlinde:2010hp}.
A regular way to test the duality  between the thermodynamic and gravitational approaches
is provided by evaluating the same observables in the state of equilibrium and in an external gravitational
field \cite{Prokhorov:2019yft, Becattini:2017ljh}.

While the chiral magnetic effect is related \cite{Kharzeev:2022hqz} to the gauge chiral anomaly, it was
suggested \cite{Landsteiner:2011cp, Jensen:2012kj, Stone:2018zel} that
the gravitational chiral anomaly \cite{Alvarez-Gaume:1984}
\begin{eqnarray}
\nabla_{\mu} j_A^{\mu} = \mathscr{N}{\epsilon}^{\mu\nu\alpha\beta} R_{\mu\nu\lambda\rho}{R_{\alpha\beta}}^{\lambda\rho}\,,
 \label{anom_N}
\end{eqnarray}
can be invoked to predict the thermal chiral vortical
effect (CVE) $ {\bf j}_A  \sim \mathscr{N} T^2 {\bf \Omega}$, where $ j_A^{\mu} $ is the axial current, $ {\bf \Omega} $ is the angular velocity of the fluid and $ T $ is the temperature \footnote{Although the
relation suggested has been verified (see, e.g. \cite{Landsteiner:2011cp, Jensen:2012kj, Stone:2018zel}) for
constituents with spin 1/2 there is an apparent mismatch
between evaluation of the thermal vortical effect within
thermal field theory and the use of gravitational anomaly \cite{Prokhorov:2020okl, Prokhorov:2021bbv, Huang:2018aly} for higher spins.}. In (\ref{anom_N}) $ R_{\mu\nu\lambda\rho} $ is the Riemann curvature tensor, $ {\epsilon}^{\mu\nu\alpha\beta}=\frac{1}{\sqrt{-g}} \varepsilon^{\mu\nu\alpha\beta}$ is the Levi-Civita symbol in curved space-time, $ \nabla_{\mu} $ is the covariant derivative, and $ \mathscr{N} $ is a numerical factor. 

In this paper we construct a hydrodynamic gradient expansion valid up to the third order in the presence of an external gravitational field (see also \cite{Romatschke:2009kr, Diles:2019uft}). If we neglect, for the moment, the terms explicitly containing the gravitational field, then the axial current takes on the form
\begin{eqnarray} \label{KVE}
j_{\mu}^{A} = \lambda_1 (\omega_{\nu}\omega^{\nu})\omega_{\mu} + \lambda_2 (a_{\nu}a^{\nu})w_{\mu}\,,
\end{eqnarray}
where $\lambda_1,\, \lambda_2$ are dimensionless constants, $ \omega_{\mu}=\frac{1}{2} {\epsilon}_{\mu\nu\alpha\beta}u^{\nu}\partial^{\alpha}u^{\beta}$ is the vorticity, and $a_{\mu}=u^{\nu}\partial_{\nu}u_{\mu}$ is the acceleration of the fluid flow.

We show, that there is a relation between $(\lambda_1 - \lambda_2)$ and the factor $\mathscr{N}$
in front of the gravitational chiral anomaly (\ref{anom_N})
\begin{eqnarray}
\frac{\lambda_{1}-\lambda_{2}}{32}&=& \mathscr{N}
\,, \label{sol2}
\end{eqnarray}
so the current (\ref{KVE}) is induced by the gravitational chiral anomaly.

The third-order current (\ref{KVE}) does not depend explicitly on the medium parameters $ T $ and $ \mu $ being a purely kinematical vortical effect (KVE). Despite its relationship with the gravitational anomaly, the KVE survives in the usual four-dimensional flat space-time. This situation is analogous to the case with the CVE and the gauge axial anomaly \cite{Son:2009tf, Zakharov:2012vv}\footnote{Obviously, considering an accelerated and/or rotating fluid in flat spacetime, even passing to a non-inertial rest frame of the fluid, we will have $ R_{\mu\nu\alpha\beta} = 0 $.}.

Eq. (\ref{sol2}) is verified by direct comparison of the transport coefficients with the anomaly factor for the Dirac field.
We also show that using the relationship between the acceleration and the thermal radiation temperature from the Unruh effect \cite{Unruh:1976db, Prokhorov:2019yft}, one can obtain an analogue of the thermal current proportional to the anomaly (cf. \cite{Landsteiner:2011cp, Jensen:2012kj, Stone:2018zel}).

We use the system of units $ e=\hbar=c=k_B=1 $, and the signature $ \eta_{\mu\nu}=(1,-1,-1,-1) $.

%================
\section{Cubic terms in gradients from the gravitational chiral anomaly}
\label{sec_main}
%================

Let us consider an uncharged non-dissipative fluid of massless fermions with an arbitrary spin in an external gravitational field with the metric $ g_{\mu\nu} (x)$. This fluid moves with a four-velocity $ u_{\mu}(x) $ and has a proper temperature $ T(x) $.

Hydrodynamic effects associated with quantum anomalies can be derived by considering the second law of thermodynamics for the entropy flow \cite{Son:2009tf}. It was recently shown in \cite{Yang:2022ksq} (see also \cite{Buzzegoli:2020ycf}) that for the non-dissipative fluid in the global equilibrium \cite{DeGroot:1980dk, Becattini:2016stj}, it suffices to take into account only the current conservation equation. In this way in \cite{Son:2009tf, Yang:2022ksq} the relationship of the CVE-current $ {\bf j}_A \sim k\, \mu^2  {\bf \Omega}$ and the chiral gauge anomaly $ \partial j_A \sim k \,{\bf E}\cdot{\bf B} $ was substantiated. Below, we generalize \cite{Son:2009tf, Yang:2022ksq} to the case of gravitational fields and the gravitational chiral anomaly.

At the quantum level, the axial current conservation is violated due to the gravitational chiral anomaly (\ref{anom_N}). Since the anomaly has the fourth order in gradients, the terms of the third order in the hydrodynamic expansion for the current should generate it.

We will consider the system in the state of global thermodynamic equilibrium \cite{DeGroot:1980dk, Becattini:2016stj}, for which the inverse temperature vector $ \beta_{\mu} =u_{\mu}/T $ satisfies the Killing equation
\begin{eqnarray}
\nabla_{\mu} \beta_{\nu} +
\nabla_{\nu} \beta_{\mu} = 0\,,
\label{killing}
\end{eqnarray}
which means we are working with the beta frame \cite{Becattini:2014yxa}.

Due to (\ref{killing}), we obtain for the second order covariant derivative
\begin{eqnarray}
\nabla_{\mu} \nabla_{\nu} \beta_{\alpha} = {R^{\rho}}_{\mu\nu\alpha} \beta_{\rho}\,.
\label{killing1}
\end{eqnarray}
An antisymmetric combination of covariant derivatives forms the thermal vorticity tensor \cite{Buzzegoli:2017cqy}
\begin{eqnarray}
\varpi_{\mu\nu} = -\frac{1}{2} (\nabla_{\mu} \beta_{\nu} - \nabla_{\nu} \beta_{\mu})\,,
\label{vort_tensor}
\end{eqnarray}
which has one vector and one pseudovector component, corresponding to the (``thermal'') acceleration $ \alpha_{\mu} $ and vorticity $ w_{\mu} $
\begin{eqnarray}
&&\alpha_{\mu} = \varpi_{\mu\nu} u^{\nu}\,,\quad
w_{\mu} = -\frac{1}{2} {\epsilon}_{\mu\nu\alpha\beta} u^{\nu} \varpi^{\alpha \beta}\,, \nonumber \\
&&\varpi_{\mu\nu}= {\epsilon}_{\mu\nu\alpha\beta} w^{\alpha} u^{\beta}+\alpha_{\mu}u_{\nu}-\alpha_{\nu}u_{\mu}\,.
\label{decomp_vort_inv}
\end{eqnarray}
In the state of global equilibrium (\ref{killing}), $ w_{\mu} $ and $ \alpha_{\mu} $ are proportional to kinematic vorticity $ \omega_{\mu}$ and acceleration $ a_{\mu}$
\begin{eqnarray}
	\alpha_{\mu} = \frac{a_{\mu}}{T}\,, \quad
	w_{\mu} = \frac{\omega_{\mu}}{T}\,.
	\label{kinemat}
\end{eqnarray}

Similarly, the Riemann tensor can be decomposed into 2 symmetric tensors and one nonsymmetric traceless pseudotensor
\begin{eqnarray}
	A_{\mu\nu} &=& u^{\alpha}u^{\beta} R_{\alpha\mu\beta\nu}\,,\quad
	B_{\mu\nu} = \frac{1}{2} {\epsilon}_{\alpha\mu\eta\rho} u^{\alpha}u^{\beta} {R_{\beta\nu}}^{\eta\rho}\,, \nonumber \\
	C_{\mu\nu} &=& \frac{1}{4} {\epsilon}_{\alpha\mu\eta\rho} {\epsilon}_{\beta\nu\lambda\gamma} u^{\alpha}u^{\beta} R^{\eta\rho\lambda\gamma}
	 \,,
	 \label{decomp_curv}
\end{eqnarray}
which are a covariant generalization of three-dimensional tensors from \cite{Landau:1975pou}. These tensors have the properties similar to the three-dimensional ones
\begin{eqnarray}
&&A_{\mu\nu} = A_{\nu\mu}\,,\quad C_{\mu\nu} = C_{\nu\mu}\,,\quad
{B^{\mu}}_{\mu}=0\,, \nonumber \\
&&A_{\mu\nu} u^{\nu} = C_{\mu\nu} u^{\nu} = B_{\mu\nu} u^{\nu}= B_{\nu\mu} u^{\nu} =0\,.
\label{ABC_prop}
\end{eqnarray}
The inverse formula has the form
\begin{eqnarray}
R_{\mu\nu\alpha\beta} &=& u_{\mu}u_{\alpha} A_{\nu\beta}+
u_{\nu}u_{\beta} A_{\mu\alpha}-u_{\nu}u_{\alpha} A_{\mu\beta}-u_{\mu}u_{\beta} A_{\nu\alpha} \nonumber \\
&&+{\epsilon}_{\mu\nu\lambda\rho}u^{\rho}(u_{\alpha} {B^{\lambda}}_{\beta}-
u_{\beta} {B^{\lambda}}_{\alpha})\nonumber \\
&&+ {\epsilon}_{\alpha\beta\lambda\rho}u^{\rho}(u_{\mu} {B^{\lambda}}_{\nu}-u_{\nu} {B^{\lambda}}_{\mu})\nonumber \\
&&+ {\epsilon}_{\mu\nu\lambda\rho} {\epsilon}_{\alpha\beta\eta\sigma} u^{\rho} u^{\sigma} C^{\lambda\eta}
\,.
\label{decomp_curv_inv}
\end{eqnarray}
The expansion (\ref{decomp_curv_inv}) is similar to the expansion of the thermal vorticity tensor (\ref{decomp_vort_inv}) or the electromagnetic field tensor.

Using (\ref{decomp_curv_inv}), one can rewrite formulas with curvature in terms of tensors $ A_{\mu\nu},\,B_{\mu\nu},\, C_{\mu\nu} $, in particular, for some scalars and pseudoscalars we obtain
\begin{eqnarray}
&&{\epsilon}^{\mu\nu\alpha\beta} R_{\mu\nu\lambda\rho}{R_{\alpha\beta}}^{\lambda\rho} = 16 (A_{\mu\nu}-C_{\mu\nu})B^{\mu\nu}\,, \nonumber \\
&&R = 2 (A^{\mu}_{\mu}-C^{\mu}_{\mu})\,,\quad
u^{\mu}u^{\nu}R_{\mu\nu} = A^{\mu}_{\mu} \,.
\label{scalar}
\end{eqnarray}

We will neglect the back reaction of matter to the gravitational field, considered as external, which allows us to impose an additional condition on the field
\begin{eqnarray}
R_{\mu\nu}=0\,.
\label{assump}
\end{eqnarray}
Although this condition is not mandatory, it eliminates 10  degrees of freedom that do not contribute to the gravitational chiral anomaly (\ref{anom_N}), which can be expressed through the Weyl tensor. Also it can be considered as an analogue of the condition of the constancy of the electromagnetic field $ F_{\mu\nu} $ in \cite{Yang:2022ksq}. 

Taking into account (\ref{assump}) we will have additional properties (compare with \cite{Landau:1975pou})
\begin{eqnarray}
A_{\mu\nu}=-C_{\mu\nu}\,,\quad A_{\mu}^{\mu}=0\,,\quad B_{\mu\nu}=B_{\nu\mu}\,,
\label{ABC_prop_1}
\end{eqnarray}
and thus the gravitational field has 10 independent components.

The contribution to the axial current is expressed in terms of all possible pseudovectors arising in the third order in gradients (compare with \cite{Romatschke:2009kr, Diles:2019uft})
\begin{eqnarray}
j_{\mu}^{A(3)} &=& \xi_{1}(T) w^2 w_{\mu} + \xi_{2}(T) \alpha^2 w_{\mu}  + \xi_{3}(T) (\alpha w) w_{\mu} \nonumber \\
&& + \xi_{4}(T) A_{\mu\nu}w^{\nu}+ \xi_{5}(T) B_{\mu\nu}a^{\nu}\,.
\label{current_decomp}
\end{eqnarray}
The unknown coefficients $ \xi_n(T) $ depend on the proper temperature $ T $.
The absence of other terms in the expansion (\ref{current_decomp}) follows from (\ref{ABC_prop_1}) and the Bianchi identity.

Substituting (\ref{current_decomp}) into (\ref{anom_N}), we thus obtain
\begin{eqnarray}
\nabla_{\mu}j^{\mu}_{A(3)} &=& (\alpha w) w^2 \left(-3 T \xi_{1}+T^2 \xi_{1}' + 2 T \xi_{3}\right) \nonumber \\
&& + (\alpha w) \alpha^2 \left(-3 T \xi_{2} + T^2 \xi_{2}' - T \xi_{3}+ T^2 \xi_{3}'\right)\nonumber \\
&& + A_{\mu\nu} \alpha^{\mu} w^{\nu} \left(T^2 \xi_{4}'+3 T \xi_{5} + 2 T^{-1} \xi_{2}+ T^{-1} \xi_{3}\right)\nonumber \\
&& + B_{\mu\nu} w^{\mu} w^{\nu} \left(-2T^{-1} \xi_{1} - 3 T \xi_{4} - T \xi_{5}\right)\nonumber \\
&& + B_{\mu\nu} \alpha^{\mu} \alpha^{\nu}\left(T^{2} \xi_{5}' - T \xi_{5} -T^{-1} \xi_{3} \right)\nonumber \\
&& + A_{\mu\nu} B^{\mu\nu}\left(-T^{-1} \xi_4 + T^{-1} \xi_{5}\right) \nonumber \\
& =&32 \mathscr{N}  A_{\mu\nu} B^{\mu\nu}
\,.
\label{main}
\end{eqnarray}
When differentiating, we used the equations, following from (\ref{killing}) and (\ref{killing1})
\begin{eqnarray}
\nabla_{\mu} T &=& T^2 \alpha_{\mu}\,,\nonumber \\
\nabla_{\mu} u_{\nu} &=& T ({\epsilon}_{\mu\nu\alpha\beta} u^{\alpha} w^{\beta}+u_{\mu}\alpha_{\nu})\,,\nonumber \\
\nabla_{\mu} w_{\nu} &=& T (-g_{\mu\nu}(w\alpha)+\alpha_{\mu}w_{\nu})-T^{-1}B_{\nu\mu}\,,\nonumber \\
\nabla_{\mu} {\alpha}_{\nu} &=& T \big(w^2(g_{\mu\nu}-u_{\mu}u_{\nu})-\alpha^2 u_{\mu}u_{\nu}-w_{\mu}w_{\nu} \nonumber \\
&&-u_{\mu}\eta_{\nu}-u_{\nu}\eta_{\mu}\big)+T^{-1}A_{\mu\nu}\,,\nonumber \\
\nabla^{\mu}(A_{\mu\nu} w^{\nu}) &=& -3 T B_{\mu\nu}w^{\mu}w^{\nu}-T^{-1} A_{\mu\nu} B^{\mu\nu}\,,\nonumber \\
\nabla^{\mu}(B_{\mu\nu} {\alpha}^{\nu}) &=& 3 T A_{\mu\nu}w^{\mu}{\alpha}^{\nu}+T^{-1} A_{\mu\nu}B^{\mu\nu} \nonumber \\
&& - T B_{\mu\nu}w^{\mu}w^{\nu} - T B_{\mu\nu}\alpha^{\mu}\alpha^{\nu} 	\,,
\label{diff_prop}
\end{eqnarray}
where $ \eta_{\mu} = \epsilon_{\mu\nu\rho\sigma} u^{\nu} w^{\rho} \alpha^{\sigma} $. The first of the equations corresponds to the well-known Luttinger relation \cite{Luttinger:1964zz}. Since (\ref{main}) contains independent pseudoscalars, we arrive at a system of equations for the unknown coefficients
\begin{eqnarray}
-3 T \xi_{1}+T^2 \xi_{1}' + 2 T \xi_{3} &=& 0\,, \nonumber \\
-3 T \xi_{2} + T^2 \xi_{2}' - T \xi_{3}+ T^2 \xi_{3}' &=& 0\,, \nonumber \\
T^2 \xi_{4}'+3 T \xi_{5} + 2 T^{-1} \xi_{2}+ T^{-1} \xi_{3}	&=& 0\,, \nonumber \\
-2T^{-1} \xi_{1} - 3 T \xi_{4} - T \xi_{5}	&=& 0\,, \nonumber \\
T^{2} \xi_{5}' - T \xi_{5} -T^{-1} \xi_{3} 	&=& 0\,, \nonumber \\
-T^{-1} \xi_4 + T^{-1} \xi_{5}	-32 \mathscr{N} & =&  0	\,.
\label{system_koef}
\end{eqnarray}
If the theory does not contain other dimensional parameters than temperature, then
\begin{eqnarray}
&&\xi_1 = T^3 \lambda_1\,,\quad
\xi_2 = T^3 \lambda_2\,,\quad
\xi_3 = T^3 \lambda_3\,, \nonumber \\
&&\xi_4 = T \lambda_4\,,\quad
\xi_5 = T \lambda_5\,,
\label{system_koef1}
\end{eqnarray}
where $\lambda_n$ are dimensionless constants. Since the number of unknowns in (\ref{system_koef}) exceeds the number of equations, the solution
relates the unknown coefficients
\begin{eqnarray}
&&\lambda_3 = 0\,,\quad \lambda_4 = - 8 \mathscr{N} - \frac{\lambda_{1}}{2}\,,\quad  \lambda_5 = 24 \mathscr{N} -\frac{\lambda_1}{2}\,, \nonumber \\
&&
\frac{\lambda_1-\lambda_2}{32} = \mathscr{N}\,.
\label{sol1}
\end{eqnarray}

First, it turns out that the current $j^A_{\mu}= \xi_{3} (\alpha w) w_{\mu} $ is absent. This condition was obvious in advance, since this term in the absence of gravitational field would violate the conservation of the current, as discussed in \cite{Prokhorov:2018bql}. 

Also from (\ref{sol1}) follows the relationship between the kinematic and the gravitational terms in the current (\ref{current_decomp})
\begin{eqnarray}
\lambda_{1}-\lambda_{2} = \lambda_{5} -\lambda_{4}\,.
\label{sol1a}
\end{eqnarray}

Finally, system (\ref{sol1}) contains Eq. (\ref{sol2}). It fixes the relationship between the gravitational chiral anomaly (\ref{anom_N}) and the transport coefficients in $ w^2 w_{\mu} $ and $ \alpha^2 w_{\mu}$. Passing to the flat space-time $ R_{\mu\nu\alpha\beta}=0 $ and kinematic quantities (\ref{kinemat}), we find that the axial current (\ref{current_decomp}) has the form (\ref{KVE})
and the transport coefficients are related to the anomaly (\ref{anom_N}) by the Eq. (\ref{sol2}).

Although the KVE (\ref{KVE}) looks like just a kinematic effect, it still depends on the properties of the medium, since velocity, acceleration, and vorticity characterize the fluid flow. This is illustrated, for example, by the Luttinger relation between the acceleration and the temperature gradient (\ref{diff_prop}). Moreover, at a finite mass, an explicit dependence on the properties of the medium appears in $ \lambda_1 $ and $ \lambda_2 $ as it is shown in \cite{Prokhorov:2018bql}.

%================
\section{Verification: Dirac field}
\label{compare}
%================

Formulas (\ref{sol2}) and (\ref{sol1}) can be verified directly by comparing the transport coefficients with the factor in the gravitational chiral anomaly. Let us consider a simple but important case of massless fermions with spin 1/2.

In \cite{Prokhorov:2018bql} (see also \cite{Palermo:2021hlf}), the following formula was obtained for the axial current in flat space-time for Dirac field
\begin{eqnarray}
	j_{\mu}^{A} &=& \left(\frac{T^2}{6}+\frac{\mu^2}{2 \pi^2} - \frac{\omega^2}{24 \pi^2}-\frac{a^2}{8 \pi^2}\right)\omega_{\mu}\,,
	\label{dirak_current_old}
\end{eqnarray}
expressed in terms of the kinematic quantities (\ref{kinemat}).

Eq. (\ref{dirak_current_old}) was obtained on the basis of the Zubarev density operator using quantum field theory at a finite temperature. Also it was obtained by calculating the exact trace over Fock space in \cite{Ambrus:2021eod, Ambrus:2019ayb}. The term $ \omega^3 $ was derived in the original papers \cite{Vilenkin:1979ui, Vilenkin:1980zv} \footnote{
The relationship with the gravitational anomaly resolves the ambiguity in $ \omega^3 $  mentioned in \cite{Stone:2018zel}.}.

Comparing (\ref{dirak_current_old}) and the well-known result for the gravitational anomaly of the Dirac field $ \mathscr{N} =1/{384 \pi^2} $ \cite{Alvarez-Gaume:1984}, we see that (\ref{sol2}) is fulfilled
\begin{eqnarray}
\left(-\frac{1}{24 \pi^2} + \frac{1}{8 \pi^2}\right) \Big/ 32=\frac{1}{384 \pi^2} \,.
\label{sum1}
\end{eqnarray}

Keeping also the terms with the gravitational field, using (\ref{sol1}) we will have
\begin{eqnarray}
j_{\mu}^{A(3)} &=& \left( - \frac{\omega^2}{24 \pi^2}-\frac{a^2}{8 \pi^2}\right)\omega_{\mu} +\frac{1}{12\pi^2} B_{\mu\nu} a^{\nu}\,.
\label{dirak_current_new}
\end{eqnarray}

%================
\section{KVE and Unruh effect}
\label{Unruh}
%================

It is possible to establish a non-trivial relationship between KVE and CVE, if we take into account the effects of thermal radiation in spacetime with the event horizon.

Current in (\ref{current_decomp}), taking into account (\ref{sol1}), can be decomposed into anomalous and conserved parts
\begin{eqnarray} \label{decomp}
j_{\mu}^A &=& j_{\mu (\text{cons})}^A + j_{\mu (\text{anom})}^A \,, \nonumber \\
j_{\mu (\text{cons})}^A &=& \frac{\lambda_1+\lambda_2}{2}\Big\{(\omega^2+a^2)\omega_{\mu}-\frac{1}{2}A_{\mu\nu}\omega^{\nu}-\frac{1}{2}B_{\mu\nu}a^{\nu}\Big\}\,, \nonumber \\
j_{\mu (\text{anom})}^A &=& 16 \mathscr{N}\Big\{(\omega^2-a^2)\omega_{\mu}-A_{\mu\nu}\omega^{\nu}+B_{\mu\nu}a^{\nu}\Big\}\,.
\end{eqnarray}
Note that the first term in $ j_{\mu (\text{anom})}^A $ is determined by the square of the thermal vorticity tensor, since $ \omega^2-a^2 =-\frac{T^2}{2} \varpi_{\mu\nu}\varpi^{\mu\nu} $.

On the other hand, in an accelerated frame, an analogue of the horizon of a black hole and the thermal radiation associated with it, called the Unruh radiation, arise \cite{Unruh:1976db, Prokhorov:2019yft}.
In the limit of a slowly rotating medium $ |\omega|\ll |a| $, the temperature of a system both with rotation and acceleration should be approximately equal to the temperature of a uniformly accelerated frame (the famous Unruh temperature) $ T_U \simeq |a|/2\pi + \mathcal{O}(\omega)$, with $ a_{\nu}a^{\nu} = -|a|^2 $.
Substituting $ |a| = 2 \pi T_U $ into $ j_{\mu (\text{anom})}^A $ for the case $ |\omega|\ll |a| $ and a flat space-time, we obtain
\begin{eqnarray} \label{new2}
j_{\mu (\text{anom})}^A \simeq 64 \pi^2\mathscr{N}\,  T_U^2 \omega_{\mu}\,.
\end{eqnarray}
Anomalous current (\ref{new2}) for the Dirac field, up to the replacement $ T_U \to T $, corresponds to the well-known CVE current
\begin{eqnarray} \label{new3}
j_{\mu (\text{anom})}^A \simeq \frac{T^2}{6} \omega_{\mu}\qquad T \leftrightarrow T_U\,,
\end{eqnarray}
and turns out to be proportional to the factor $ \mathscr{N} $ from the gravitational chiral anomaly.

In an accelerated medium with a finite temperature the mean values of physical quantities depend on the proper temperature and the proper acceleration, which are independent parameters. However, since at the Unruh temperature the system is in a Minkowski vacuum state, physical quantities vanish at $ T=T_U$ \cite{Becattini:2017ljh, Prokhorov:2019yft}. In the case with both vorticity and acceleration the situation is more complicated and a simple vacuum cancellation condition is not evident. Instead of this we obtain a CVE-like thermal vortical current. The anomalous current of the form (\ref{new2}), was also obtained in another approach \cite{Stone:2018zel}, using the condition of the cancellation at the event horizon for a metric similar to the Kerr black hole.

%================
\section{Conclusion}
\label{sec_concl}
%================

In this paper, we have demonstrated that the
gravitational chiral anomaly is imprinted in properties
of vortical and accelerated matter even in the absence of
gravitational fields. There is a relation, see Eq. (\ref{sol2}),
between the transport coefficients in the third order
in gradients and the overall factor in front of the
anomaly. Thus, one can talk about a new anomalous
transport phenomenon -- the kinematical vortical effect
(KVE), which is a kind of extension of the equivalence
principle to higher powers of acceleration. Although the
derivation given is valid only in the limit of dissipation-free fluid and massless constituents with arbitrary spin, the universality of
the gravity suggests a possibility of generalizing the
results obtained to other systems. This point deserves
further consideration.

The magnitude of the effect reflects the strength of
the coupling of spin of the constituents to gravitational
field. In case of the Dirac field the relation obtained agrees
with the known transport coefficients as evaluated in
the limit of non-interacting gas, see discussion around
Eq. (\ref{sum1}). Also, using the decomposition into anomalous and conserved parts, we obtain a thermal vortical current related to the Unruh temperature and induced by the gravitational anomaly.

%In case of spin-3/2 constituents we have used the Rarita-Scwinger-Adler model
%\cite{Adler:2017shl}
%to perform calculation of the relevant transport coefficients.
%The results agree with predictions made on the basis
%of gravitational anomaly and incorporate, in particular,
%strong spin dependence inherent to the anomaly, see Eq.
%(\ref{sum19}).

On the academic side, the results obtained
provide further explicit examples of equivalence
between statistical and gravitational approaches, see
Introduction. From the point of view of the experiment,
KVE could provide a unique opportunity to search for
the manifestations of the gravitational anomaly outside
the physics of a curved space-time.

%=================================================
{\bf Acknowledgements}
%=================================================

The authors are thankful to M. Buzzegoli, S. Ghosh, Z. V. Khaidukov,  P. G. Mitkin, V. P. Nair, A. V. Sadofyev, and G. Torrieri for stimulating discussions and interest in the work. GP is grateful to V. V. Srabionyan for help with computer calculations. The work was supported by Russian Science Foundation Grant No. 21-12-00237, the work of VIZ is partially supported by grant No. 0657-2020-0015 of the Ministry of Science and Higher Education of Russia.

\bibliography{lit}

\end{document}